# Kalman filter/deep-learning hybrid automatic boundary tracking of optical coherence tomography data for deep anterior lamellar keratoplasty (DALK)

Hongrui Yi*[a], Jinglun Yu[a], Yaning Wang[a], Justin Opfermann[b], Bill G. Gensheimer[c,d], Axel Kriger[b], and Jin U. Kang[a]

[a] Department of Electrical and Computer Engineering, Johns Hopkins University, Baltimore, MD 21218, USA; [b] Department of Mechanical Engineering, Johns Hopkins University, Baltimore, MD 21218, USA; [c] Department of Ophthalmology, White River Junction VA Medical Center, White River Junction, VT 05001, USA; [d] Department of Ophthalmology, Dartmouth-Hitchcock Medical Center, Lebanon, NH 03766, USA

## ABSTRACT

Deep anterior lamellar keratoplasty (DALK) is a highly challenging partial thickness cornea transplant surgery that replaces the anterior cornea above Descemet's membrane (DM) with a donor cornea. In our previous work, we proposed the design of an optical coherence tomography (OCT) sensor integrated needle to acquire real-time M-mode images to provide depth feedback during OCT-guided needle insertion during Big Bubble DALK procedures. Machine learning and deep learning techniques were applied to M-mode images to automatically identify the DM in OCT M-scan data. However, such segmentation methods often produce inconsistent or jagged segmentation of the DM which reduces the model accuracy. Here we present a Kalman filter based OCT M-scan boundary tracking algorithm in addition to AI-based precise needle guidance to improve automatic DM segmentation for OCT-guided DALK procedures. By using the Kalman filter, the proposed method generates a smoother layer segmentation result from OCT M-mode images for more accurate tracking of the DM layer and epithelium. Initial *ex vivo* testing demonstrates that the proposed approach significantly increases the segmentation accuracy compared to conventional methods without the Kalman filter. Our proposed model can provide more consistent and precise depth sensing results, which has great potential to improve surgical safety and ultimately contributes to better patient outcomes.

## 1. INTRODUCTION

The DALK is a partial thickness corneal transplant procedure that offers significant advantages over full-thickness transplants, including reduced risk of graft rejection and better preservation of endothelial cell density [1]. The DALK technique specifically replaces the anterior stroma above DM while leaving the endothelium intact, making it a preferred option for treating conditions such as keratoconus, stromal scarring, and other corneal pathologies [2]. However, the success of this procedure hinges on the precise separation of the deep stroma from DM, a process that is technically challenging due to the delicate nature of the tissue and the micron-scale accuracy required.

Further author information: (Send correspondence to Hongrui Yi)
*Hongrui Yi: E-mail: hyi13@jhu.edu; Phone: +1 513-599-9487

The Big Bubble technique [3] is a widely adopted method for DALK, where a needle is inserted into the corneal stroma to inject air and pneumodissect the layers. Accurate placement of the needle is critical to avoid perforation of the DM, which would necessitate conversion to a full-thickness transplant, negating the benefits of DALK. The OCT has emerged as a transformative imaging modality in ophthalmic surgery [4][5], providing high-resolution, cross-sectional images of corneal layers. By integrating an OCT sensor with the needle, our previous work demonstrated the potential of real-time depth feedback to guide needle insertion during DALK procedures. Machine learning models, particularly deep learning approaches like U-Net, were employed to segment corneal layers in OCT M-scan data, enabling autonomous or semi-autonomous needle guidance. However, these methods are often challenged by signal noise, motion artifacts, and data inconsistencies, which can lead to jagged or inaccurate segmentation of the DM and epithelium boundary as shown in Figure 1 deep learning approaches like U-Net based approach , were employed to segment corneal layers in OCT M-scan data, enabling autonomous or semi-autonomous needle guidance. However, these methods are often challenged by signal

To address these challenges, we propose a novel Kalman filter/deep-learning hybrid (KDH) automatic boundary tracking approach that combines deep learning-based segmentation with a Kalman filter. The Kalman filter's predictive capabilities smooth out segmentation artifacts and enhance the robustness of boundary tracking, making it particularly suited for noisy and dynamic surgical environments [6][7]. By integrating this method into the OCT-guided DALK workflow, the system improved the consistency and accuracy of DM and epithelium segmentation, ultimately enhancing the safety and efficacy of the procedure. Experiments demonstrate significant improvements in segmentation accuracy and robustness, highlighting the potential of this KDH approach to advance the state of OCT-guided microsurgery.

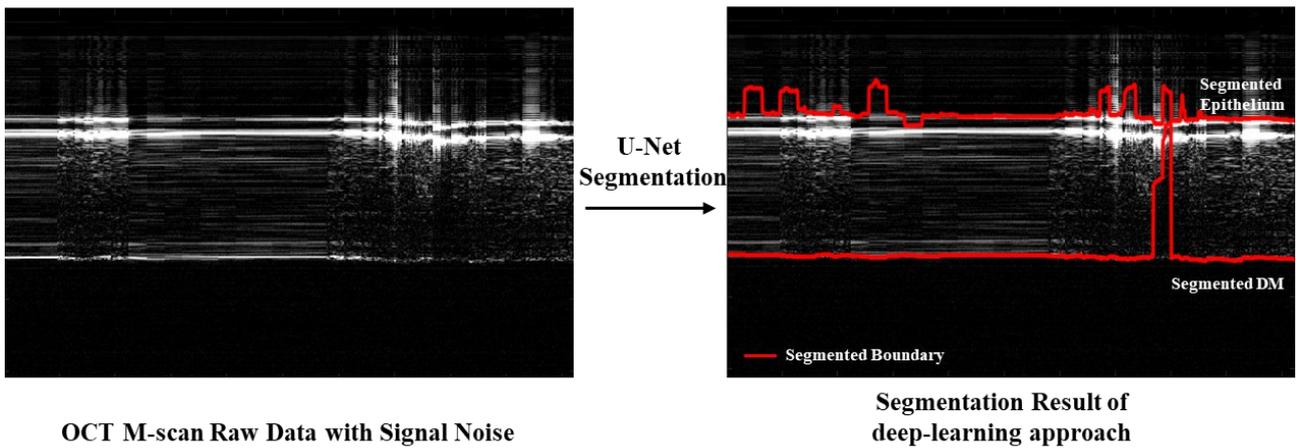

Figure 1. Example of the inconsistent or jagged segmentation of the DM and epithelium

## 2. METHODS

### 2.1 Data Collection

The dataset for this study consisted of *ex vivo* corneal samples collected from 12 rabbit eyes [8][9], acquired using a custom-built OCT imaging system. Each sample was subjected to a Big Bubble DALK procedure to simulate clinical conditions. M-mode OCT images were captured during the procedure, focusing on the stromal layers and Descemet's membrane. A total of 250 OCT images were collected, of which 200 were used for training and 50 for testing. Ground truth labels for DM and epithelium boundaries were manually annotated by experienced ophthalmologists [1], ensuring accuracy and reliability of the dataset. Preprocessing steps included image normalization, denoising, and augmentation to enhance segmentation robustness.

### 2.2 Deep learning Based Segmentation

The deep learning-based segmentation approach used a U-Net [5] architecture as shown in Figure 2 (a), which is widely regarded for its effectiveness in biomedical image segmentation. The network consists of a contracting path and an expansive path, enabling it to capture contextual information while maintaining spatial resolution. Input OCT M-scan images were preprocessed with normalization and denoising techniques to enhance the signal-to-noise ratio (SNR). The U-Net's encoder-decoder structure, combined with skip connections, facilitated precise localization of boundaries within the corneal tissue.

Training was performed on 200 annotated *ex vivo* M-scans using a 20% cross-validation split to minimize overfitting. The loss function used cross-entropy loss coefficient to handle class imbalance and ensure accurate boundary segmentation. Despite its ability to detect DM and epithelium boundaries, the U-Net occasionally produced jagged and inconsistent segmentations due to signal noise and fluctuation in the OCT data. This limitation motivated the integration of the Kalman filter in the following steps to smooth the segmentation results and improve tracking consistency.

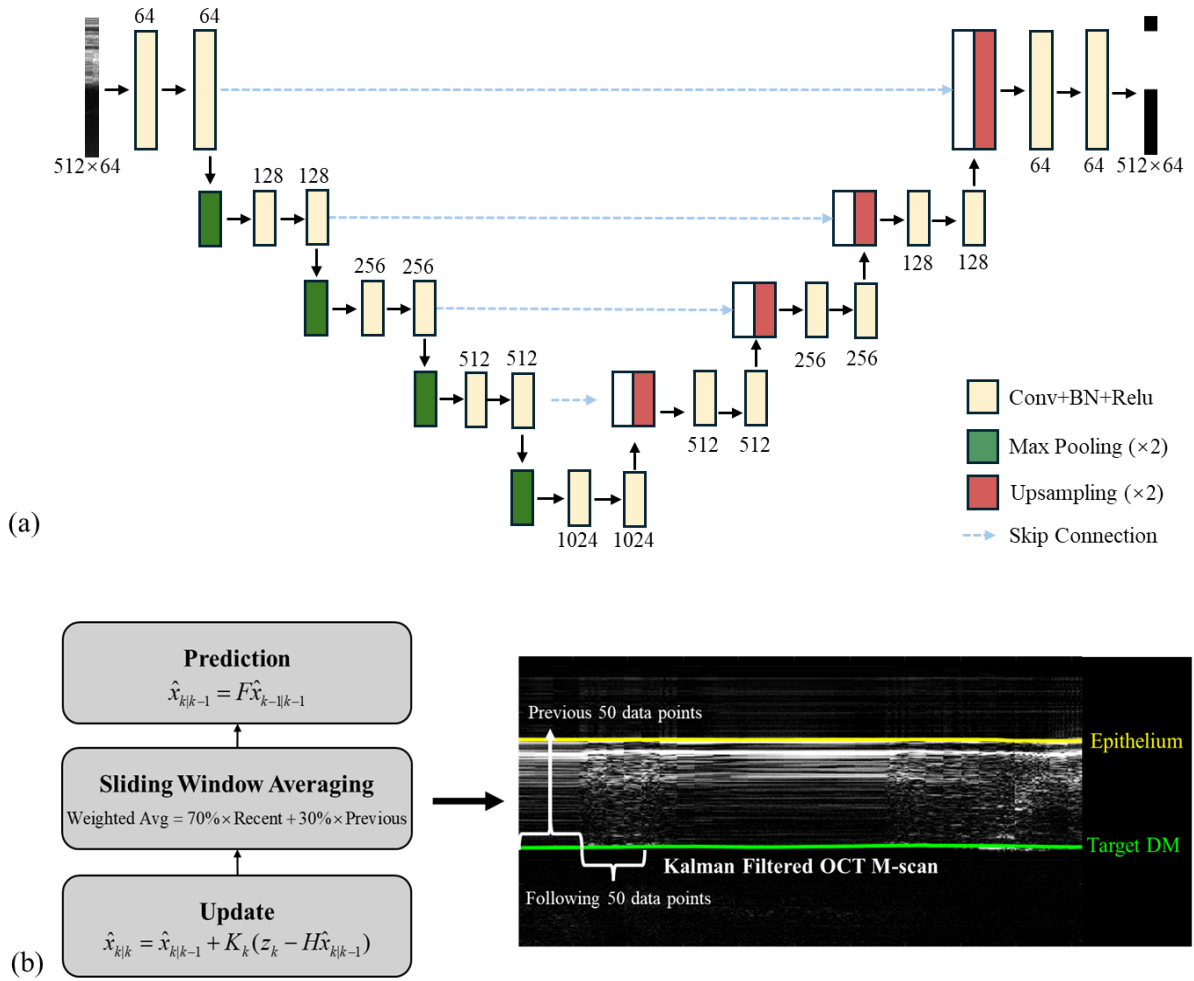

Figure 2. (a) Modified U-Net architecture (b) Illustration of Kalman Filter Integration

**2.3 Kalman Filter Integration**

To enhance the accuracy and robustness of corneal layers' segmentation, we integrated the Kalman filter into the workflow to address the limitations of purely deep learning-based methods as shown in Figure 2 (b). The Kalman filter was configured with a state transition matrix $F = 1$, an observation matrix $H = 1$, a process noise covariance $Q = 1 \times 10^{-5}$, and an observation noise covariance $R = 1$. This configuration provided a well-balanced tradeoff between the smoothing effect and the responsiveness of the filter. The prediction and update formulas are defined as:

$$\hat{x}_{k|k-1} = F\hat{x}_{k-1|k-1} \tag{1}$$

$$\hat{x}_{k|k} = \hat{x}_{k|k-1} + K_k(z_k - H\hat{x}_{k|k-1}) \tag{2}$$

A sliding window technique was employed to refine the observation inputs further. For the initial 50 data points, the standard Kalman update was applied to establish a reliable state estimate. As additional data accumulated, the observation input was adaptively adjusted using a weighted average of the most recent 50 data points with 70% weight and the preceding 50 data points with 30% weight. This adaptive approach as shown in Figure 2 (b) ensured responsiveness to new data while maintaining stability, resulting in an ameliorate and consistent line for the corneal layers.

During segmentation, the Kalman filter utilized predictions from the deep learning model as inputs, dynamically refining the positions of the DM and epithelium boundaries in real-time. The integration of the Kalman filter significantly minimize noise and artifacts, ensuring smooth and accurate boundary tracking even under dynamic surgical conditions.

## 3. EXPERIMENTS AND RESULTS

### 3.1 Data Description

In the experiments, the *ex vivo* dataset was divided into 200 image pairs for training and 50 image pairs for testing. To accommodate the requirements of M-mode OCT data and real-time tracking, each 512 × 512 image pair was automatically cropped into patches of size 16 × 512 × 32 for input into the network. These patches were normalized using the mean and standard deviation computed from their respective training sets. During inference, the network reconstructed the processed patches back into their original 512 × 512 format, ensuring output integrity for accurate analysis. The trained model was then employed for real-time segmentation of the *ex vivo* M-scan, with real-time recordings of both the deep learning-based segmentation results and the KDH segmentation results, which combined deep learning and the Kalman filter.

### 3.2 Performance Evaluation

The performance of the KDH segmentation results combining deep learning and the Kalman filter was compared to the only deep learning-based segmentation results. All experiments were conducted on an NVIDIA GeForce RTX 4070 Ti SUPER GPU. The evaluation emphasized robustness under signal noise, motion artifacts, and data inconsistencies, segmentation accuracy for tracking corneal layers (DM and epithelium), and inference time efficiency to assess the potential for real-time guidance

The segmentation performance of the epithelium is illustrated in Figure 3 (a) In the M-scan OCT data for *ex vivo*-1, where the signal quality is high, the filtered data aligns closely with the ground truth, exhibiting no prediction offset. However, in *ex vivo*-2, where the signal strength is lower, the segmentation result that only based on deep learning shows noticeable fluctuations, whereas the proposed KDH segmentation method maintains smooth and stable results. In cases involving motion artifacts and signal loss, as seen in *ex vivo*-3 and *ex vivo*-4, the deep learning-based segmentation produces inconsistent and jagged boundary lines. In contrast, our KDH approach demonstrates superior robustness, delivering consistent and stable segmentation results even under challenging data collection conditions. The segmentation performance for the DM is presented in Figure 3 (b). In the case of stable signal conditions as shown in *ex vivo*-1, the segmentation results show no significant deviations, with both the deep learning-based and KDH methods achieving accurate results. However, when signal loss occurs in *ex vivo*-2, the segmentation results only based on deep-learning exhibits noticeable fluctuations, while proposed KDH method maintains a smooth and stable segmentation result. In the case of jagged segmentation as shown in *ex vivo*-3 and *ex vivo*-4, the deep learning-based results become increasingly inconsistent, producing irregular and fragmented boundaries. This issue is further magnified in cases with severe artifacts, where the conventional deep learning segmentation shows significantly more jagged edges. In contrast, our KDH approach consistently provides robust and stable segmentation results, demonstrating its ability to handle even the most challenging conditions effectively.

To further validate the advantages of our proposed method, we evaluated the tracking accuracy of the KDH approach against the conventional deep learning-based method, as shown in Table 1 The average absolute errors are reported in both pixels and actual distances (µm), with each pixel in the M-mode data corresponding to 2.61 µm. The results indicate that our method achieves a 59.55% reduction in the average epithelium error and a 48.15% reduction in the average DM error compared to the conventional method. These notable improvements demonstrate the stability and accuracy of the KDH approach in tracking critical corneal boundaries.

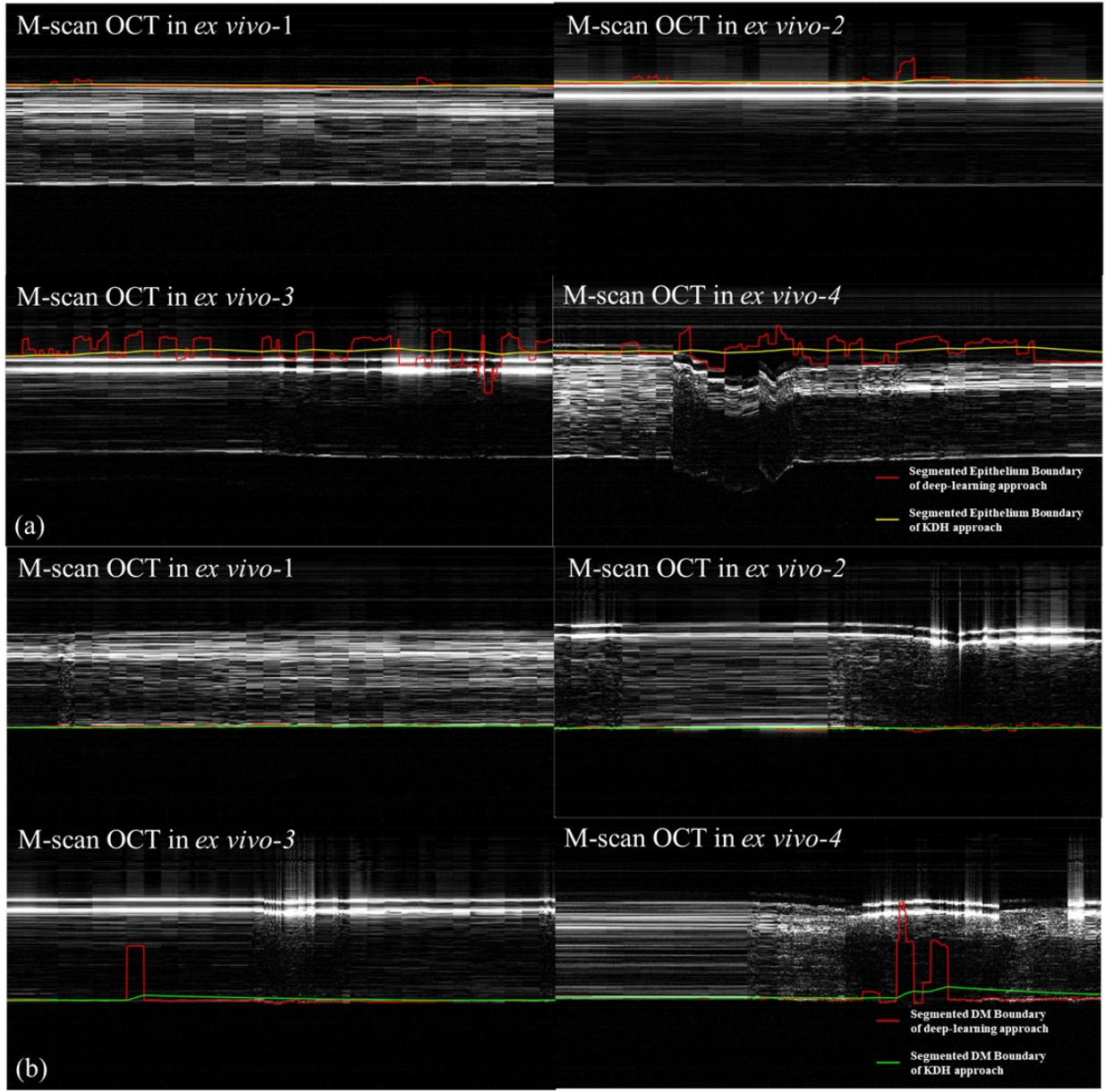

Figure 3. Comparison of segmentation results between the KDH approach and the deep-learning approach. (a) Epithelium segmentation result comparison (b) DM segmentation result comparison

Table 1. Comparison of segmentation results between the Kalman-filter/deep-learning KDH approach and the deep-learning (DL) approach

| Approach | Average Epithelium Error | | Average DM Error | |
|---|---|---|---|---|
| KDH | 0.36 pixel | 0.9396 μm | 0.28 pixel | 0.7308 μm |
| Deep learning-based | 0.89 pixel | 2.3229 μm | 0.54 pixel | 1.4094 μm |

We further evaluated the proposed model by applying it to segment data acquired from our OCT-based, sensor-integrated, eye-mounted robotic system [4] during a real-time robotic-assisted DALK procedure. A sample video (Figure 4) demonstrates the process, showcasing the model's ability to accurately and seamlessly segment the epithelium and DM throughout the robotic needle insertion. These findings highlight the potential of our approach to deliver reliable and robust real-time guidance for DALK procedures.

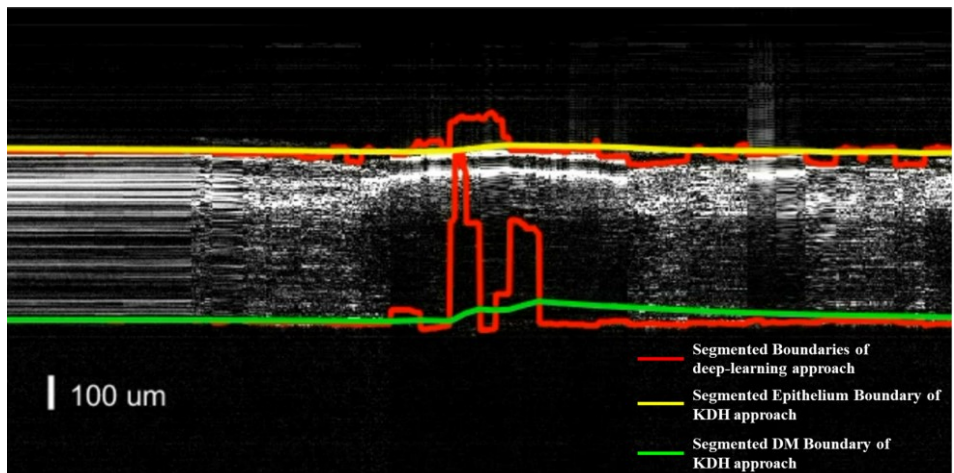

Figure 4. Video 1-Sample video of real-time segmentation and tracking of DM and epithelium layers using deep-learning and KDH methods http://dx.doi.org/doi.number.goes.here

## 4. DISCUSSION AND CONCLUSION

In this study, we introduced a Kalman filter/deep-learning hybrid automatic boundary tracking approach for automatic boundary tracking in OCT-guided DALK procedures. By leveraging the Kalman filter's predictive smoothing capabilities, the proposed method addressed challenges associated with noisy and inconsistent OCT data, significantly enhancing segmentation accuracy and robustness. Our results underscore the potential of this KDH approach to improve the safety and efficacy of DALK, paving the way for further advancements in autonomous ophthalmic surgery.

## ACKNOWLEDGEMENTS

This research was supported by the National Institute of Biomedical Imaging and Bioengineering of the National Institutes of Health under award number 1R01EY032127 (PI: Jin U. Kang). The study was conducted at Johns Hopkins University and Dartmouth-Hitchcock Medical Center.